\documentclass[prd,twocolumn,groupedaddress,showpacs,nofootinbib]{revtex4}
\pdfoutput=1
\usepackage{graphicx}
\usepackage{bm}
\usepackage{times}
\usepackage{slashed}
\usepackage{color}
\usepackage{color}

\begin{document}

\title{Dilepton bounds on left-right symmetry at the LHC run II and neutrinoless double beta decay}
\author{Manfred Lindner, Farinaldo S.\ Queiroz, Werner Rodejohann}

\affiliation{$^{1}$Max-Planck-Institut f\"ur Kernphysik, Postfach 103980, 69029 Heidelberg, Germany \\
}
\begin{abstract}
\noindent 
In the light of the new 13 TeV dilepton data set with $ 3.2\, {\rm fb^{-1}}$ integrated luminosity from the ATLAS collaboration, we derive limits on the $Z^{\prime}$ mass in the context 
of left-right symmetric models and exploit the complementarity with dijet and $lljj$ data, 
as well as neutrinoless double beta decay. We keep the ratio of the left- and right-handed gauge coupling  
free in order to take into account different patterns of left-right symmetry breaking.  By combining the dielectron and dimuon data we can exclude $Z^{\prime}$ masses below $3$~TeV for $g_R=g_L$, and for $g_R \sim 1$ we rule out masses up to $\sim 4$~TeV. Those comprise the strongest direct bounds on the $Z^{\prime}$ mass from left-right models up to date. We show that in the usual plane of right-handed neutrino and charged gauge boson mass, dilepton data can probe a region of parameter space inaccessible to neutrinoless double beta decay and $lljj$ studies.  Lastly, through the mass relation between $W_R$ and $Z^{\prime}$ we present an indirect bound on the lifetime of neutrinoless double beta decay using dilepton data. Our results prove that the often ignored dilepton data in the context of left-right models actually provide important complementary limits.

\end{abstract}

\maketitle

\noindent
\section{Introduction} 
Left-Right (LR) symmetric models based on the gauge group 
$SU(3)_C \otimes SU(2)_L\otimes SU(2)_R\otimes U(1)_{B-L}$ gauge symmetry \cite{Pati:1974yy,Mohapatra:1974hk,Mohapatra:1974gc,Senjanovic:1975rk,Senjanovic:1978ev,Mohapatra:1979ia} 
are compelling extensions of the 
Standard Model (SM), in particular because 
they address parity violation at the weak scale and 
active neutrino masses via the seesaw mechanism. 
Albeit, the scale at which parity is restored is not predicted, leaving room for a large range of 
gauge boson masses ($Z_R$ and $W_R^\pm$) which set the 
scale of symmetry breaking.  
In 
general, the main collider search strategies for the gauge bosons 
are based on dilepton, diboson, and dilepton plus dijet data. 
Besides collider searches for those gauge bosons a multitude of studies in the context of meson \cite{Beall:1981ze,Zhang:2007da,Bertolini:2014sua}, flavor \cite{Castillo-Felisola:2015bha,Dekens:2014ina,Das:2012ii} and neutrinoless double beta decay data \cite{Hirsch:1996qw,Tello:2010am,Nemevsek:2011aa,Barry:2013xxa,Huang:2013kma,Awasthi:2013ff,Chakrabortty:2012mh,Dev:2013vxa}
have set important limits in TeV scale LR symmetric models.

Those studies often assume that the left- and right-handed gauge couplings are identical, 
i.e.\  $g_R=g_L$, and that the charged gauge boson is lighter than the neutral one. In  particular, in the context of minimal left-right symmetric models this relation reads $M_{Z^{\prime}} = 1.7 M_{W_R}$ for $g_R=g_L$. Since in principle, those gauge bosons share similar production cross sections at the LHC, the best way to constrain a LR symmetry is by performing $W_R$ searches in the light of the above mass relation. Another motivation is the fact that the $W_R$ mass is straightforwardly connected to the $SU(2)_R$ gauge coupling and left-right symmetry breaking, $M_{W_R} = g_R v_R$.

However, $W_R$ searches based on $lljj$ studies lose sensitivity for sufficiently heavy right-handed neutrinos, $M_N \gtrsim M_{W_R}$ \cite{Keung:1983uu}. Moreover, there are many ways to successfully break the left-right symmetry down to the Standard Model yielding either $g_R \neq g_L$ or $M_{Z^{\prime}} \ll M_{W_R}$, or both \cite{Chakrabortty:2016wkl}.

Therefore, in this work we remain agnostic as to how precisely the left-right symmetry is broken and 
perform an  independent  $Z^{\prime}$ search with LHC constraints and show that actually the use of dilepton data from the LHC offers an interesting avenue to probe left-right models. Indeed, as long as neutral gauge bosons couplings to 
charged leptons are not suppressed, neutral gauge boson searches based on dilepton data are rather promising 
and give rise to the most stringent limits on their masses \cite{Alvares:2012qv,Cogollo:2012ek,Frandsen:2012rk,Alves:2013tqa,Profumo:2013sca,Arcadi:2013qia,Dudas:2013sia,Dong:2014wsa,Martinez:2014ova,Arcadi:2014lta,Alves:2015mua,
Kelso:2014qka,Richard:2014vfa,Lebedev:2014bba,Kahlhoefer:2015bea,Okada:2016gsh,
Allanach:2015gkd,Alves:2015pea,Rodejohann:2015lca,Gupta:2015lfa,Arcadi:2015nea,
Guella:2016dwo,Queiroz:2016awc}. 

Along this line, using 8 TeV centre-of-mass energy, ATLAS and CMS collaborations with integrated luminosity $\mathcal{L} = 20-21\mbox{ fb}^{-1}$ \cite{Aad:2014cka,Khachatryan:2014fba} in the dilepton channel have found no evidence for new resonances, and consequently 95\% confidence level lower limits on the mass of the sequential and other $Z^{\prime}$ bosons were placed. In the context of left-right models recent limits based on the 8 TeV data were derived in \cite{Patra:2015bga}. 
Most recently both ATLAS and CMS collaborations have presented their results based on run II with $13$~TeV centre-of-mass energy and $2-3$ fb$^{-1}$ of integrated luminosity in \cite{ATLAS1}. In what follows we will use ATLAS results since they have more luminosity. Using $\mathcal{L}=3.2$ fb$^{-1}$ of data, the 
collaboration performed resonance searches for dilepton invariant masses above 500 GeV, and used its invariant mass spectrum as the discriminating variable. No statistical fluctuation above SM expectations has been found and 
95\% C.L.\ bounds were obtained, ruling out the sequential $Z^{\prime}$ with masses below $3.4$~TeV. In the light of this new data set we update existing limits on the $Z^{\prime}$ gauge boson of left-right models.  

As mentioned above, a popular and often considered process to constrain left-right symmetry is neutrinoless double beta decay \cite{Rodejohann:2011mu}. Here the most straightforward approach is to consider the right-handed analog of the standard light neutrino exchange mechanism, which is sensitive to the $W_R$ mass and the  right-handed neutrino masses. The LHC analog of this diagram is the production of two like-sign leptons and two jets, $eejj$. Several papers have studied the LHC constraints \cite{Khachatryan:2014dka} of this final state and obtained the corresponding limits on the parameter space in comparison to the double beta decay 
constraints \cite{Tello:2010am,Nemevsek:2011aa,Barry:2013xxa,Huang:2013kma,Awasthi:2013ff,Chakrabortty:2012mh,Dev:2013vxa}. We point out here that 
$Z'$ mass limits can within many LR models 
be translated into $W_R$ mass limits, which therefore provides indirect limits on the parameter space relevant for double beta decay. These limits are moreover essentially independent on the right-handed neutrino mass and complementary to other limits. Depending on the 
breaking scheme of the left-right symmetry and the ratio of gauge boson couplings, they probe
part of parameter space outside the one reachable 
by LHC $eejj$ searches and double beta decay, and are  stronger than the ones from dijet data.\\ 

We start this letter by discussing some aspects of 
left-right symmetry before providing the $Z'$ mass limits 
by newest LHC dilepton data and then making the comparison to direct 
limits on the double beta decay parameter space. 


\section{Left-Right Symmetry\label{sec:mod}}

Left-right models rely on the gauge group $SU(3)_C \otimes SU(2)_L \otimes SU(2)_R \otimes U(1)_{B-L}$ and have quite interesting features: (i) they naturally incorporate baryon and lepton number; (ii) generate neutrino masses through type I+II seesaw mechanisms; (iii) might appear in unification theories such as SO(10) and E(6); (iv) they restore C and/or P (charge conjugation and parity) at high energy scales, thus addressing their violation at the electroweak scale, which is arguably the most striking motivation for a gauge left-right symmetry. Within this context we address two different exemplary symmetry breaking patterns. 

\subsection{Scalar Content A}

In case there is a left-right discrete symmetry in the model, the fields transform under
parity and charge conjugation as follows: $P$: $Q_L \leftrightarrow Q_R, \phi \leftrightarrow  \phi^{\dagger}, \Delta_{L,R} \leftrightarrow  \Delta_{R,L}$; and $C$: $Q_L \leftrightarrow Q_R^c, \phi \leftrightarrow  \phi^{T}, \Delta_{L,R} \leftrightarrow  \Delta^{\ast}_{R,L}$. The full fermion and scalar content of the model is 
\begin{eqnarray}
Q_L & = & \left (
\begin{array}{c}
u_L \\
d_L 
\end{array}
\right ), Q_R  =  \left (
\begin{array}{c}
u_R \\
d_R 
\end{array}
\right ) ,\nonumber\\
 l_L & = & \left (
\begin{array}{c}
\nu_L \\
e_L 
\end{array}
\right ),  l_R  =  \left (
\begin{array}{c}
N_R \\
e_R 
\end{array}
\right ),
\label{L}
\end{eqnarray}
\begin{eqnarray}
\phi & = & \left (
\begin{array}{cc}
\phi_1^0 & \phi_1^+\\
\phi_2^- & \phi_2^0
\end{array}
\right ), \Delta_{L,R}  =  \left (
\begin{array}{cc}
\delta^+_{L,R}/\sqrt{2} & \delta^{++}_{L,R} \\
\delta^0_{L,R} & -\delta^+_{L,R}/\sqrt{2}
\end{array}
\right ). \nonumber\\
\end{eqnarray}
Notice that in this most often considered left-right model 
$\phi$ is a bidoublet which transform as $(2,2^\ast,0)$ under $SU(2)_L\otimes SU(2)_R\otimes U(1)_{B-L}$ in order to generate fermion masses, and $\Delta_{L,R}$ are scalar triplets with $B-L=2$ \cite{Minkowski:1977sc,Mohapatra:1980yp,Chang:1984uy}. 

While typically $g_L = g_R$ is assumed as a consequence of a discrete left-right symmetry such as parity or charge conjugation, this is actually not necessary. 
For instance, by using so-called D-parity instead, which is broken by the vev of a total gauge singlet  field, one can easily depart from $g_L = g_R$ at low energies, see  \cite{Chang:1984uy,Awasthi:2013ff,Borah:2013lva,Deppisch:2014qpa,Patra:2014goa,Deppisch:2014zta,Patra:2015bga} for explicit realizations. 
In short, one decouples in such theories the breaking of the discrete and 
gauge left-right symmetries, consequently  
the left- and right-handed scalar fields have different masses early on, 
and the gauge couplings run differently resulting in $g_L \neq g_R$ at 
low scales. 

The relevant aspect of this class of 
models for what follows is the fact that it induces $Z^{\prime}$-fermions couplings 
\begin{eqnarray} \label{eq:coupl}
\frac{g_R}{\sqrt{1-\delta \tan^2\theta_W}}
\overline{f}\,\gamma_\mu \left(
     g_V^f - g_A^f \gamma^5 \right) \,f\,\, Z^{\prime \mu} \,,
\end{eqnarray}
with the couplings determined by 
\begin{eqnarray}
&&\hspace*{-0.5cm}g^f_V=\frac{1}{2}\left[\big\{\delta \tan^2\theta_W \left(T^{f}_{3L} 
     -\mbox{Q}^f \right)\big\} 
+ \big\{T^{f}_{3R}- \delta \tan^2\theta_W \mbox{Q}^f\big\}\right] \nonumber \\
&&\hspace*{-0.5cm}g^f_A=\frac{1}{2}\left[\big\{\delta \tan^2\theta_W \left(T^{f}_{3L} 
     -\mbox{Q}^f \right)\big\} 
- \big\{T^{f}_{3R}- \delta \tan^2\theta_W \mbox{Q}^f\big\}\right] \nonumber
\end{eqnarray} where $T^{f}_{3L,3R}= \pm 1/2$  
for $^{\rm up}_{\rm down}$-fermions,  $\delta=g^2_L/g^2_R$, and $Q^f$ being the corresponding electric charge. In general the neutral current depends on 
how the left-right gauge symmetry is broken. Therefore, 
the $Z^{\prime}$-fermion coupling strength is subject 
to the spontaneous symmetry breaking pattern. We will 
get back to that further. Another interesting outcome 
of this class of models is the mass relation
\begin{eqnarray}
\frac{M_{Z_{R}}}{M_{W_{R}}}=\frac{\sqrt{2} g_R/g_L}{\sqrt{(g_R/g_L)^2-\tan^2\theta_W}}\,.
\label{MZMW}
\end{eqnarray}
Setting $g_L= g_R$, we find $M_{Z_{R}}\simeq 1.7 M_{W_{R}}$, with $M_{W_R} =g_R v_R$, where $v_R$ is the scale at which the left-right gauge symmetry is broken to the Standard Model. 
This mass relation has profound implications, since it clearly shows that bounds on the mass of the charged gauge boson imply stronger ones on the $Z^{\prime}$ mass. Thus, as far as constraining the gauge boson masses are concerned, $W_R$ searches should be the primary target. Although, collider searches for $W_R$ rely  mainly on dilepton plus dijet data that are subject to a relatively large background, since the collaborations typically do not enforce LNV in the event selection. Hence, it is worthwhile performing an independent collider study of the $Z^{\prime}$ gauge boson in left-right models because it predicts a clean signal based on dilepton data. Moreover, in case a signal consistent with a $W_R$ is observed at the LHC a corresponding dilepton excess can be expected using the mass relation Eq.\ (\ref{MZMW}).

\subsection{Scalar Content B}

In general terms, as already in the model treated above, $g_R$ may be different from $g_L$ at low energy scale. In addition, the mass relation in Eq.\ (\ref{MZMW}) depends also on the spontaneous symmetry breaking pattern. It has been proposed in \cite{Chang:1983fu,Chang:1984uy}, that if one evokes a D-parity breaking with the following scalar particle content,
\begin{eqnarray}
\phi & = & \left (
\begin{array}{cc}
\phi_1^0 & \phi_1^+\\
\phi_2^- & \phi_2^0
\end{array}
\right ), \,\, \Delta_{L,R}  =  \left (
\begin{array}{cc}
\delta^+_{L,R}/\sqrt{2} & \delta^{++}_{L,R} \\
\delta^0_{L,R} & -\delta^+_{L,R}/\sqrt{2}
\end{array}
\right ) \nonumber\\
\Omega_{L,R} &  = & \left (
\begin{array}{cc}
\omega^0 & \omega^+/\sqrt{2}\\
\omega^-/\sqrt{2} & -\omega^0
\end{array}\right), 
\end{eqnarray}
(the fermion sector is unchanged from the previous model) one can actually get $M_{W_R} \gg M_{Z^{\prime}}$ and $g_R \neq g_L$.  
The two additional triplet scalars $\Omega_{L,R}$, which transform with $B-L=0$, are required in order to break the LR symmetry in two steps: (i) $SU(2)_L \otimes SU(2)_R \otimes U(1)_{B-L} \rightarrow \, SU(2)_L \otimes U(1)_R \otimes U(1)_{B-L}$ through the vev of neutral field of the triplet $\Omega_{R}$ which sets the $W_R$ mass; (ii) $U(1)_R \otimes U(1)_{B-L} \rightarrow U(1)_Y$, by the vev of neutral field of the triplet $\Delta_{R}$ determining the $Z_R$ mass. Its coupling to fermions is the same as in Eq.\ (\ref{eq:coupl}). 
The additional scalar fields above are needed to break the SM to electromagnetism as usual. Taking the vev of $\Omega_R$ to be at very high energy scales, the masses of the charged and neutral gauge bosons are uncorrelated, with $M_{W_R} \gg M_{Z_R}$. Thus, in this scenario, collider searches for neutral gauge bosons are more promising since the charged gauge boson is not attainable at the LHC. \\

In summary, left-right models predict the existence of new neutral and charged gauge bosons and their mass relation is subject to the scalar content of the model. If the neutral gauge boson is heavier than the charged one, our results play a complementary role since a $W_R$ signal at the LHC should be accompanied by a dilepton resonance. If the neutral gauge boson is lighter than the charged one, our limits are crucial and the most restrictive direct limits on the $Z^{\prime}$ mass. Moreover, the 
left- and right-handed gauge couplings can be different from each other, and should be left as free parameter. Hence, in an attempt of remaining agnostic regarding the precise scalar content and the spontaneous symmetry breaking pattern chosen, we base our results on the $Z^{\prime}$-fermions couplings in the context of Eq.\ (\ref{eq:coupl}) and varying in particular 
the gauge coupling $g_R$. 

\section{Dilepton Limits}

Dilepton data ($ee$, $\mu\mu$) is a promising data set in the search for new physics in several theories which possess neutral gauge bosons with sizable couplings to 
leptons\footnote{ See \cite{delAguila:2010mx} for a good review of LEP-II bounds on gauge bosons. In particular, note that the LEP-II limit on our $Z^\prime$ bosons is 
667 GeV for $g_R = g_L$.}. At the LHC, in particular, the high selection efficiencies and well understood background naturally poses this channel as a golden channel since a heavy dilepton resonance is a new physics smoking gun. Up to date, the most sensitive heavy neutral gauge bosons searches were carried out by both ATLAS and CMS collaborations \cite{Aad:2014cka,Khachatryan:2014fba}. ATLAS collaboration has obtained 95\% C.L.\ limits on the sequential $Z^{\prime}$ boson at 8 TeV centre-of-mass energy with $\mathcal{L} =20$ fb$^{-1}$ ruling out masses below 2.90 TeV, combining $ee$ and $\mu\mu$ data. With the LHC run II data at 13 TeV with $\mathcal{L}=3.2$ fb$^{-1}$ unprecedented sensitivities were reached in \cite{ATLAS1}. 

 \begin{figure}[!t]
\includegraphics[width=\columnwidth]{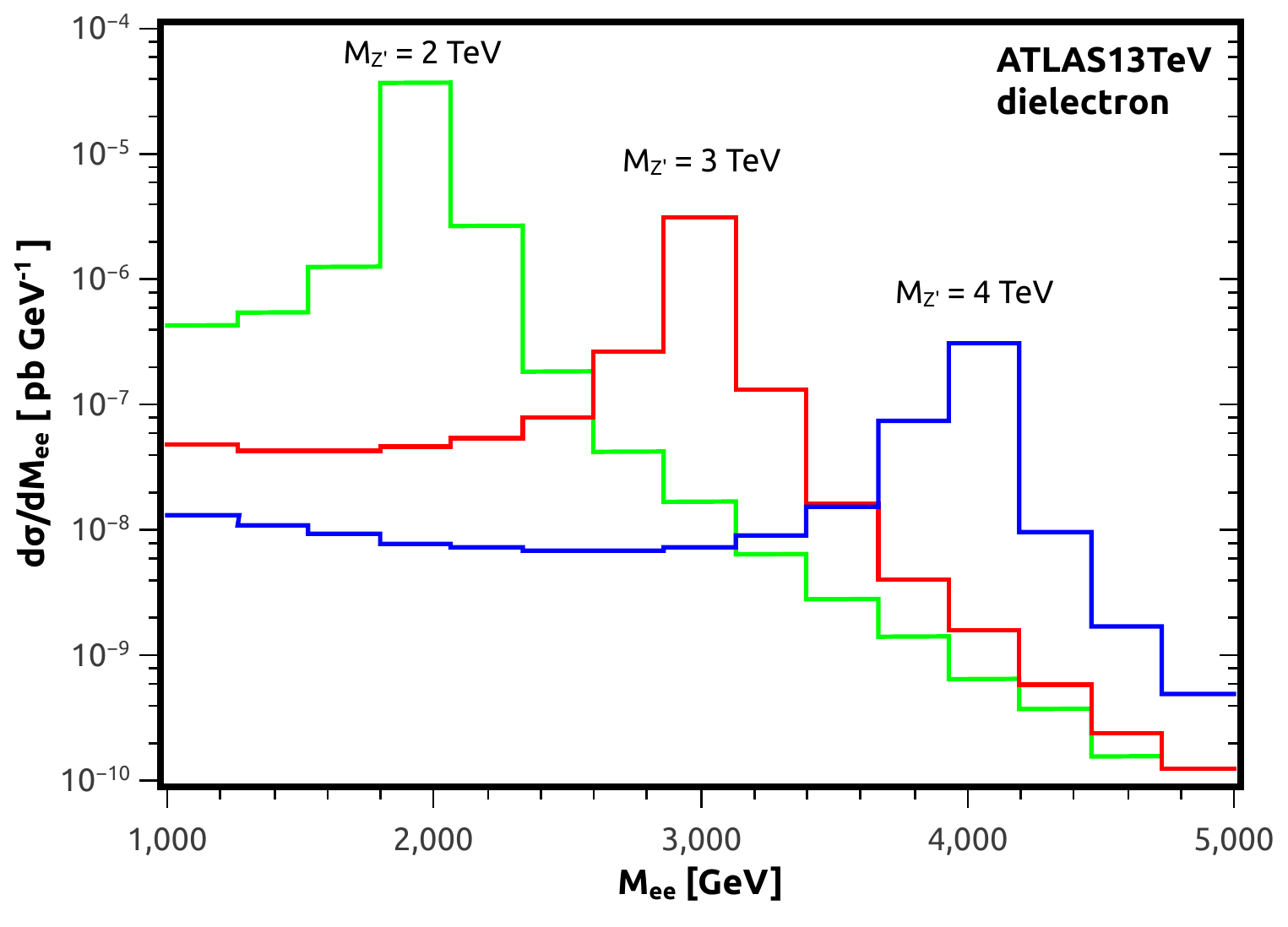}
\caption{Differential cross section for the dielectron channel at 13 TeV. The $p_T$ and $E_T$ cuts given in Sec.\ III were applied to the figure.}
\label{fig:1}
\end{figure}

The most relevant background contributions arise from the Drell-Yan processes. Additional background sources come from diboson and top-quark production. Due to misidentification of jets as electrons also known as jet-fake rate, multi-jet and $W$+jets channels are also background to dielectron searches. We have taken the background estimations from \cite{ATLAS1}. As for the signal, we simulated $pp \rightarrow Z^{\prime} \rightarrow e^+e^-,\, \mu^+ \mu^-$ plus one jet (in order to account for extra isolated dilepton events with the presence of one jet which are identified as dilepton events) using MadGraph5 \cite{Alwall:2014hca,Alwall:2011uj} for invariant masses above 500 GeV as analyzed in \cite{ATLAS1}. We have used the CTEQ6L parton distribution functions \cite{Lai:2009ne} and taken into account jet hadronization QCD radiation with Pythia and imposed the same isolation requirements as the ATLAS collaboration. As for detector effects, we adopted a flat 70\% signal efficiency,
which is reasonable as shown in Fig.\ 1 of \cite{Aad:2014cka}. The dimuon data efficiency lies around 40\%, resulting into a slight overestimation of our combined limits, but it lies within the $2\sigma$ error bar reported by ATLAS collaboration. Following the procedure in   \cite{ATLAS1}, the signal events were selected by applying the following cuts:
\begin{itemize}
\item  $E_T(e_1) > 30 \,{\rm GeV}, E_T(e_2) > {\rm 30 \,GeV}, |\eta_e| < 2.5$,
\item  $p_T(\mu_1) > 30 \,{\rm GeV}, p_T(\mu_2) > 30 \,{\rm GeV}, |\eta_{\mu}| < 2.5$,
\item $500 \,{\rm GeV} < M_{ll} < 6000 \,{\rm GeV}$,
\end{itemize}where $M_{ll}$ is the 
invariant mass of the lepton pair, which is the most important signal-to-background discriminating variable for this kind of analysis. 

\begin{figure}[!t]
\includegraphics[width=\columnwidth]{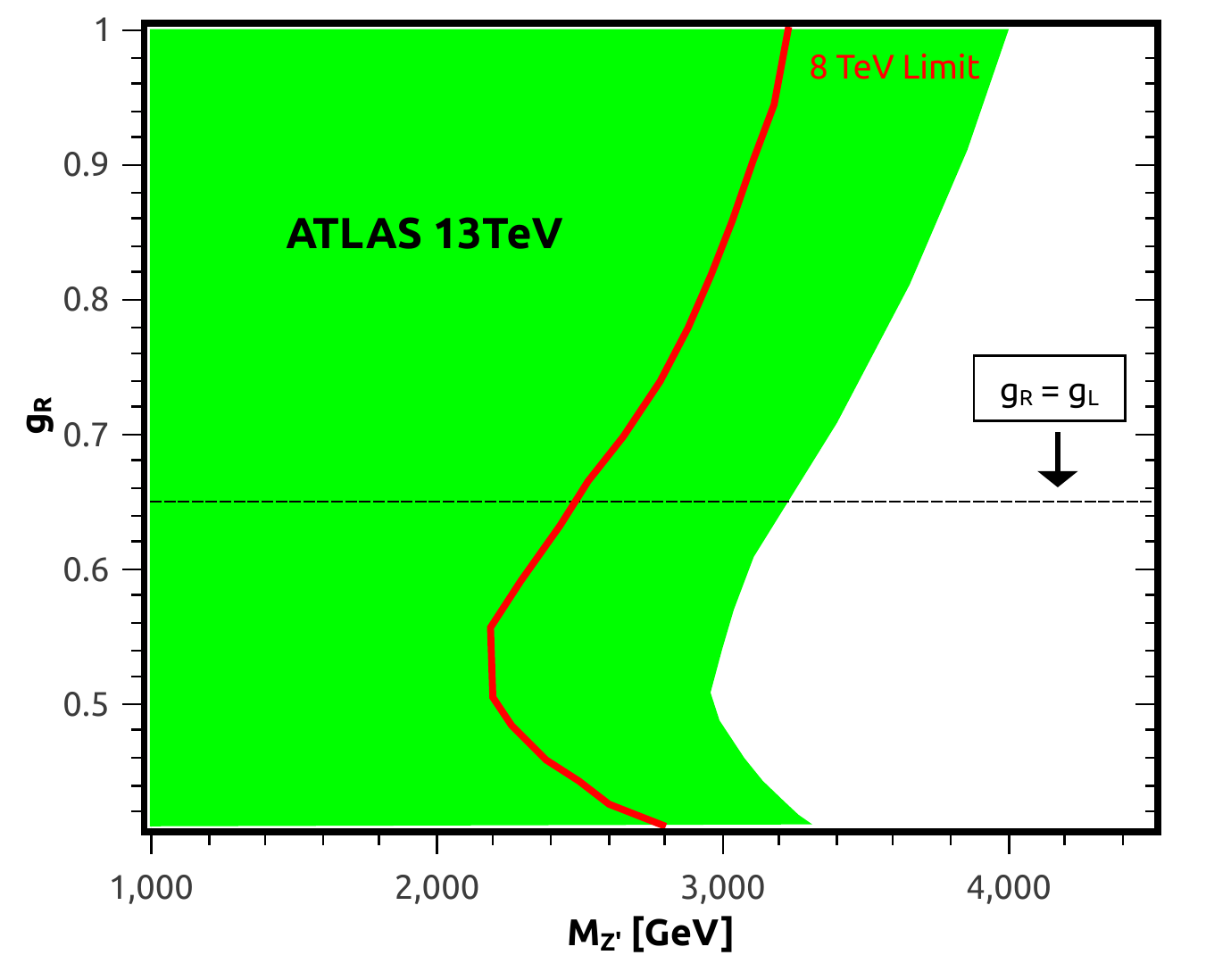}
\caption{13 TeV LHC bounds on the $Z^{\prime}$ mass in the context of left-right symmetric models for different $g_R$ values. In particular, for $g_R=g_L$ (dashed line) we find $M_{Z^{\prime}} \geq 3230$~GeV. Notice that the $Z^{\prime}$-fermion coupling strength does not always grow with $g_R$ because of the presence of extra $1/g_R^2$ factors in the vector/axial couplings, explaining the shape of the figure. Those limits are subject to $2\sigma$ errors as reported by ATLAS. }
\label{fig:3}
\end{figure}

By looking at the differential cross section in terms of the invariant mass distribution of the lepton pairs (see Fig.\ \ref{fig:1}), one can clearly see the pronounced peak when the invariant mass matches the mass of the neutral gauge boson. In the left-right models, the cross section suddenly increases near the $Z^{\prime}$ mass, featuring a narrow resonance. One could do a similar plot for the number of events and notice that there is negligible SM background for invariant masses above 2 TeV (see table 2 of \cite{ATLAS1}). Hence, in the light of no event observed for large invariant masses one can place robust limits on the $Z^{\prime}$ mass. In Fig.\ \ref{fig:3} we present the limits on the $g_R$ vs.\ $M_{Z^{\prime}}$ mass plane enforcing the signal cross section not to exceed at 95\% C.L.\ the observed one using Fig.\ 3-c of  \cite{ATLAS1}. For instance, for $g_R/g_L=1$ the limit is 
$M_{Z^{\prime}}>3230$ GeV, whereas for $g_R/g_L=1~(0.6)$ the limit is 
$M_{Z^{\prime}}> 4000~(3300)$ GeV. Using the mass relation between $M_{Z^{\prime}}$ and 
$M_{W_R}$ from Eq.\ (\ref{MZMW}), these limits translate into 
$M_{W_R}>1900$~GeV $(g_R/g_L=1)$ , $M_{W_R}> 2666$ GeV  $(g_R=1)$, and $M_{W_R}> 1100$~GeV  $(g_R/g_L=0.6)$.
 
The presence of right-handed neutrinos might degrade the limits by 1-2\% due to branching ratio subtraction into charged leptons. 
Hence, our limits are essentially independent on the right-handed neutrino mass, in contrast to limits on $W_R$ bosons from $eejj$ data, which will be of importance when we now continue to discuss the connection to double beta decay.


\section{Connection with Neutrinoless Double Beta Decay and $W_R$ searches} 

\begin{figure}[!t]
\includegraphics[width=\columnwidth]{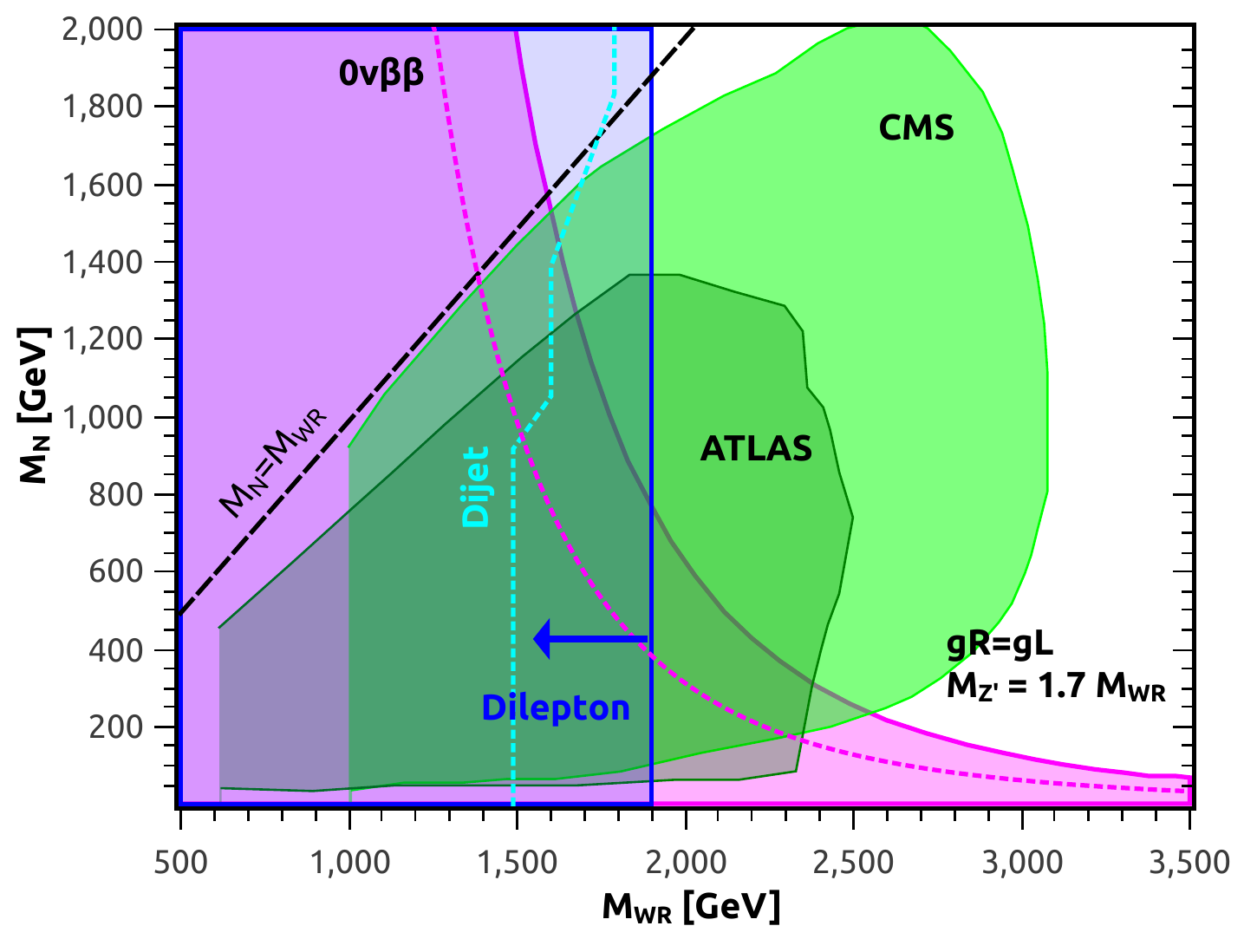}
\caption{Complementary among neutrinoless double beta decay, $W_R$ and $Z^{\prime}$ searches at the LHC for $g_R/g_L=1$.}
\label{fig:4}
\end{figure}

\begin{figure}[!t]
\includegraphics[width=\columnwidth]{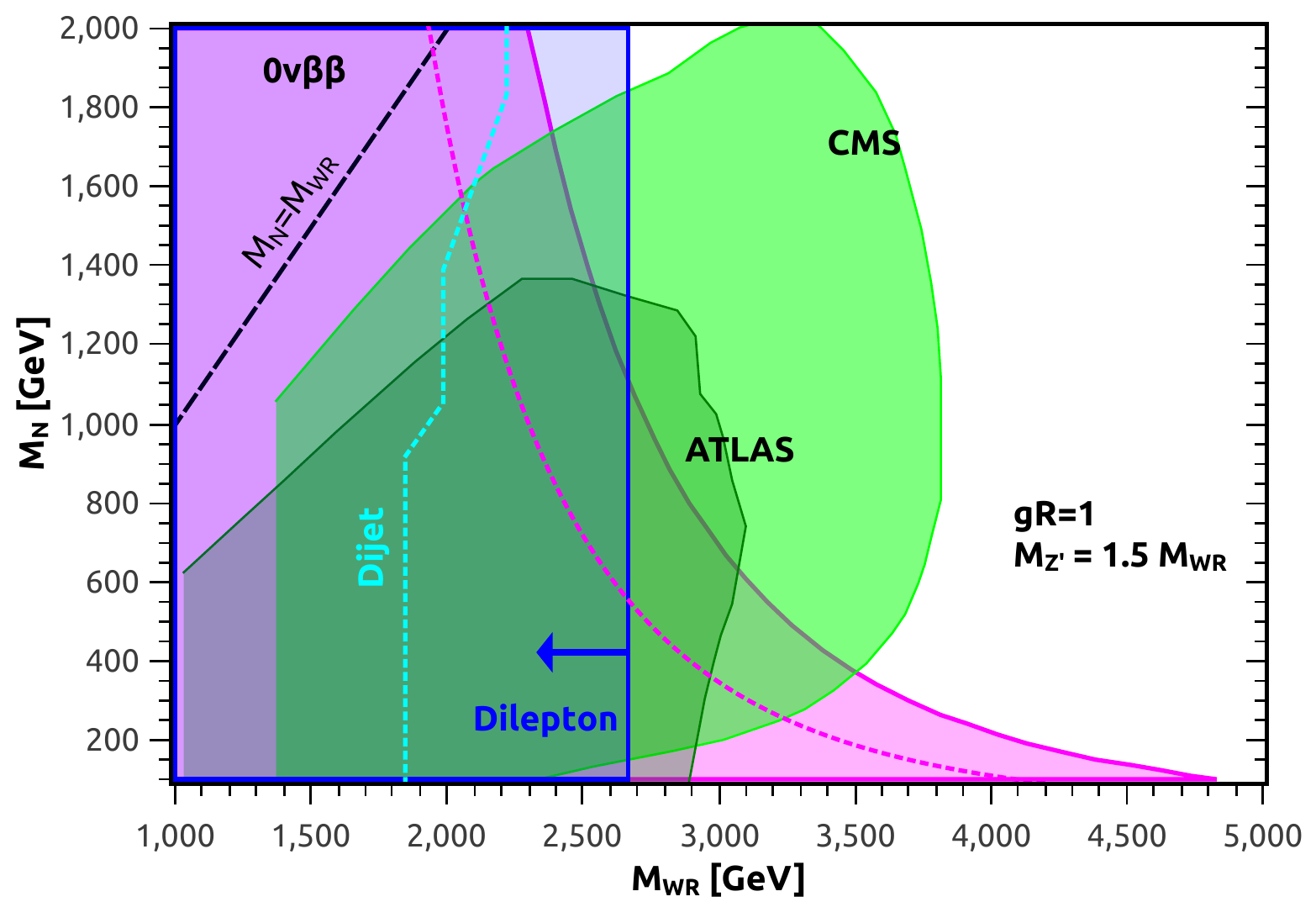}
\caption{Complementary among neutrinoless double beta decay, $W_R$ and $Z^{\prime}$ searches at the LHC for $g_R=1$, i.e. $g_R/g_L=1.54$. }
\label{fig:7}
\end{figure}

\begin{figure}[!t]
\includegraphics[width=\columnwidth]{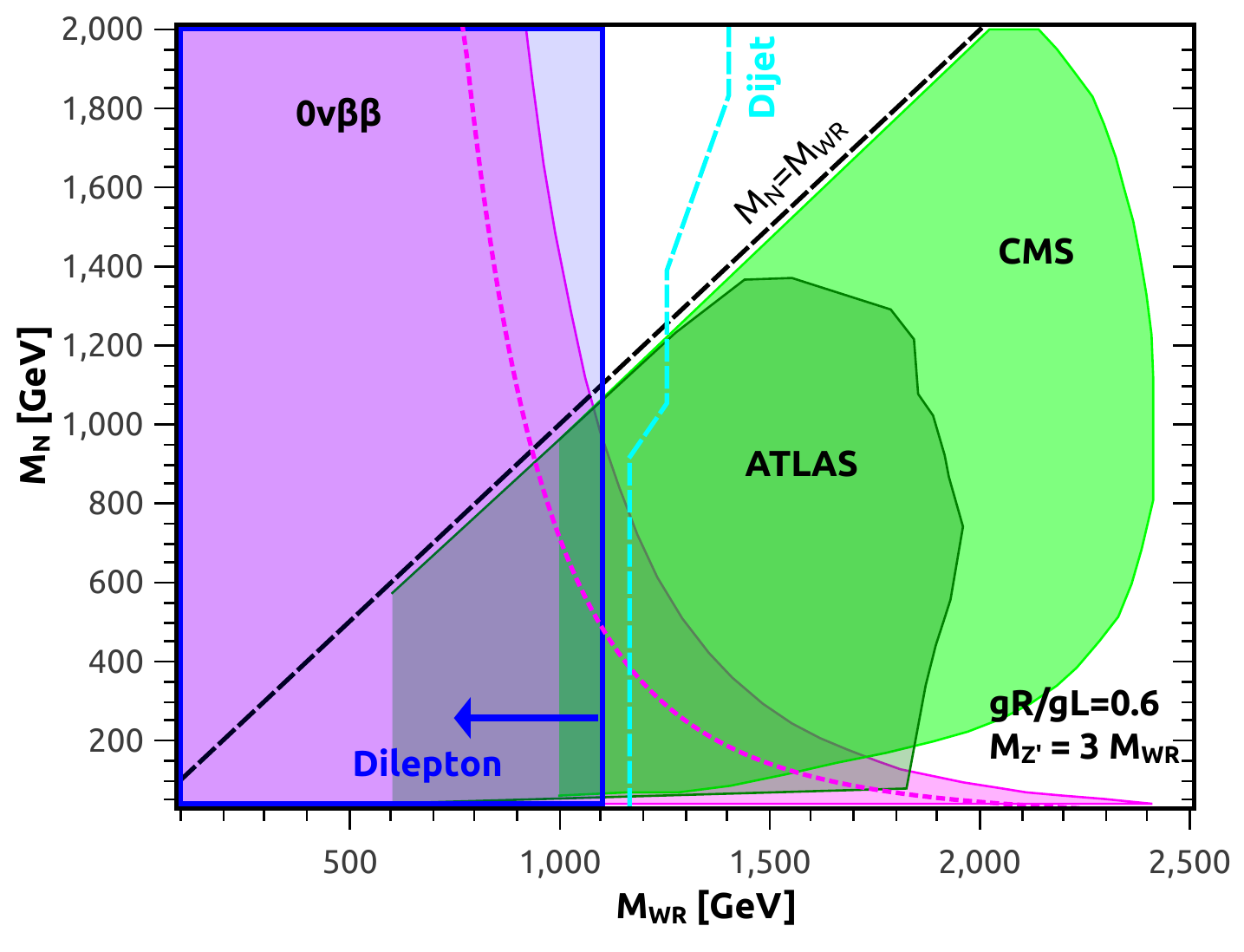}
\caption{Complementary among neutrinoless double beta decay, $W_R$ and $Z^{\prime}$ searches at the LHC for $g_R/g_L=0.6$. }
\label{fig:5}
\end{figure}

Left-right symmetric models give rise to several contributions to neutrinoless double beta decay ($0\nu\beta\beta$) \cite{Hirsch:1996qw,Rodejohann:2011mu}. Focusing on the purely right-handed neutrino exchange, one finds applying the current bound from KamLAND-Zen of $2.6\times 10^{25}$ yrs for the decay of $^{136}$Xe \cite{Asakura:2014lma} the following constraint: 
\begin{equation}
\left(\frac{g_R}{g_L}\right)^4
\frac{\left| V_{ei}^2 \right| }{M_{N_{i}} M_{W_R}^4} \leq 
0.1-0.2 \,\, {\rm TeV^{-5}}\,,
\end{equation}
where $M_{N_i}$ are the right-handed neutrino masses and $V$ is the right-handed analog of the PMNS matrix $U$, assumed here for definiteness to be equal to $U$.  Assuming the right-handed contribution to be the dominant mechanism of the decay, the pink region in Figs.\ \ref{fig:4}--\ref{fig:5} is ruled out. Also given in those figures is the region from CMS and ATLAS $eejj$ searches 
\cite{ATLAS:2012ak,Khachatryan:2014dka}, the strongest constraints 
being obtained 
for centre-of-mass energy of 8 TeV with integrated luminosity of 
$19.7$ fb$^{-1}$ from the CMS collaboration \cite{Khachatryan:2014dka} (see \cite{Gluza:2015goa,Gluza:2016qqv} for independent studies). 
No excess beyond SM expectations is observed and 95\% C.L.\ 
limits were derived ruling out masses up to $3$~TeV as shown in 
Figs.\ \ref{fig:4}--\ref{fig:5} in the $M_N$ vs.\ $M_{W_R}$ plane 
(the limits on $M_{W_R}$ are similar to the ones from meson 
physics \cite{Beall:1981ze,Zhang:2007da,Bertolini:2014sua}).
 
Note that those limits are applicable to the $M_{W_R} > M_N$ regime and assume the  branching ratio $W_R \rightarrow l N$ to be  100\%, where $l=e,\mu$. Moreover, for small $M_N \ll M_{W_R}$ the detector efficiencies are rather poor, explaining the shape in Fig.\ \ref{fig:4}. Assuming that the narrow width approximation is valid and the efficiency remains for different values of $g_R$, one can naively rescale the limits for different $g_R/g_L$ values. The branching ratio remains the same, but the production cross section goes with $g_R^2/g_L^2$. Thus, we can translate that shift in the production cross section into a rescaling of the bound on the $W_R$ as presented in Figs.\ \ref{fig:7} and \ref{fig:5}. We emphasize that the new bounds on the $W_R$ mass are approximate, and certainly overestimated in the regions which $M_{W_R} \sim M_N$ or $M_{W_R} \gg M_N$, but satisfactory to our reasoning. In Figs.\ \ref{fig:7} and \ref{fig:5}, ATLAS and CMS limits were rescaled from $g_R/g_L=1$ to $g_R = 1$ and $g_R/g_L=0.6$, respectively.  
We also include in the figure the (partly in analogy to the $eejj$ limits rescaled) 
bounds on $W_R$ as obtained by dijet 
data from Ref.\ \cite{Helo:2015ffa}\footnote{The behavior of the dijet 
limit is explained by the change in the branching 
ratio $W_R \rightarrow jj$ when $M_N > M_{W_R}$.}. For other studies regarding $W_R$ searches see \cite{Dev:2015kca,Banerjee:2015gca}.

As already mentioned, $Z^{\prime}$ searches based on dilepton data are 
essentially not sensitive to the right-handed neutrino masses. 
Therefore, relating the $Z^{\prime}$ mass limits via 
Eq.\ (\ref{MZMW}) to $W_R$ mass limits allows to set 
indirect constraints on the parameter range. This method allows 
to enter parameter space not probed by $W_R$ studies and 
neutrinoless double beta decay, as one can see in 
Figs.\ \ref{fig:4}--\ref{fig:5}. In particular, 
the bound on the $Z^{\prime}$ mass is important for 
low right-handed neutrino masses and if 
the right-handed neutrino lives in the neighborhood of 
$M_{W_R}$ (keeping in mind that vacuum stability 
prohibits the mass of the right-handed neutrino 
to be much heavier than the $W_R$ \cite{Mohapatra:1986pj,Maiezza:2016bzp}). 

Comparing dijet and dilepton data for different $g_R/g_L$ ratios, we see that 
for both $g_R=g_L$ and $g_R=1$, dilepton data surpasses the current dijet limits. 
The dilepton data might lose in sensitivity to the dijet data if there is 
a large mass splitting in the gauge boson relation $M_{Z^{\prime}} > M_{W_R}$, be it from the ratio $g_R/g_L$ or in models as described in Sec.\ IIA. 
We emphasize again that in case one breaks left-right symmetry such that the $W_R$ are much heavier 
than the $Z^\prime$ and thus disconnected, then both $jj$ and $eejj$ data yield weak  limits. 

\begin{figure}[!th]
\includegraphics[width=\columnwidth]{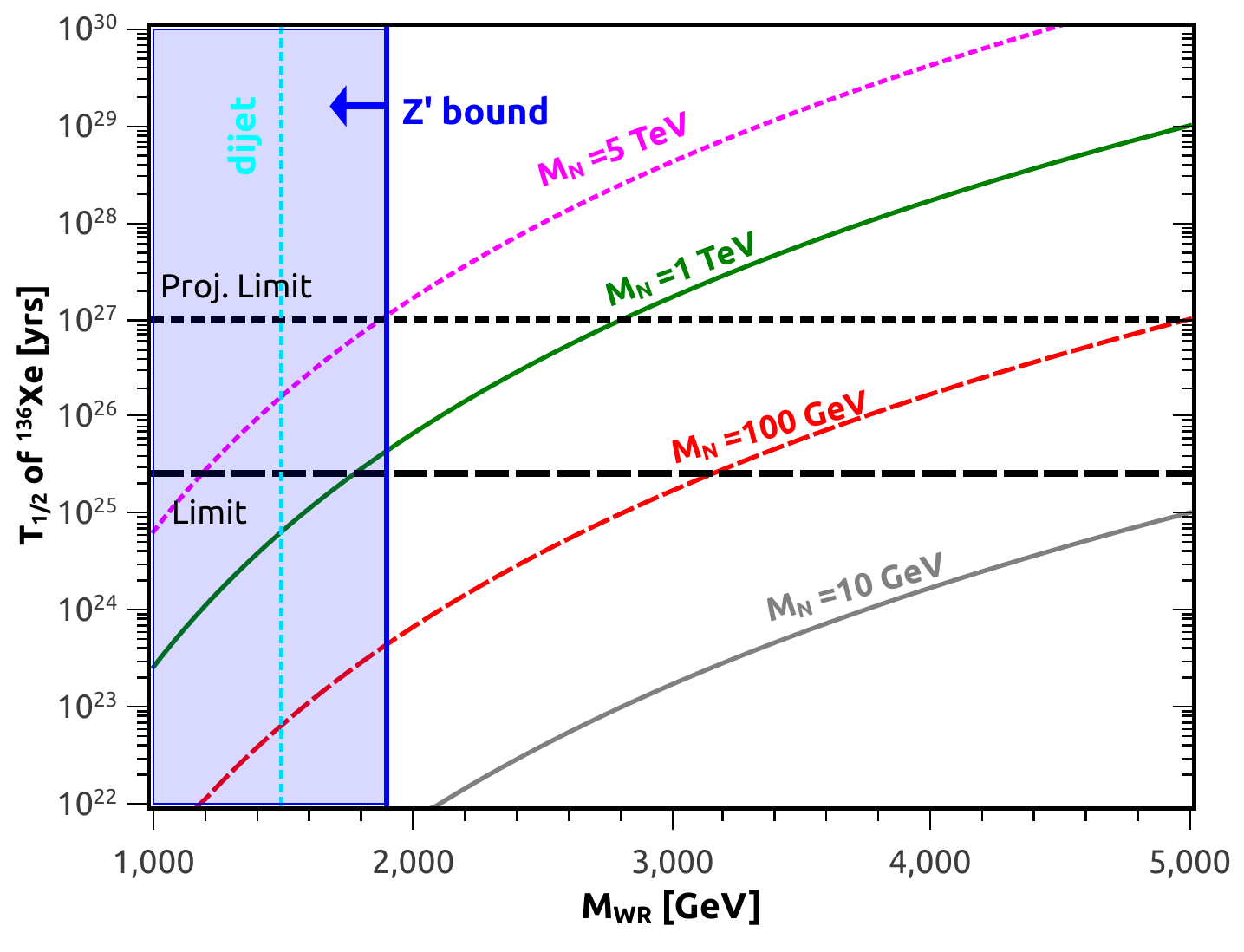}
\caption{Indirect limits on the half-life of $^{136}$Xe from dilepton and dijet searches at the LHC for $g_R/g_L=1$. In particular, for $M_N=5$~TeV, where no bound from $eejj$ searches is applicable, dilepton resonances provide the most restrictive limits on the  half-life of $^{136}$Xe.}
\label{fig:lifetime}
\end{figure}

In Fig.\ \ref{fig:lifetime} we use our bound on the $Z^{\prime}$ mass to 
constrain the half-life for double beta decay. We plot the half-life 
for $^{136}$Xe assuming different values for a right-handed neutrino mass.  
We also exhibit the dijet and current (projected) $0\nu\beta\beta$ decay 
limits. The curves from top to bottom are for 
$M_N=5$~TeV, $M_N=1$~TeV, $100$~GeV and $10$~GeV, respectively. In particular, for $M_N=5$~TeV there is no bound from $eejj$ searches on the $W_R$ mass, since $M_N > M_{W_R}$, and only those stemming from dijet and dilepton resonances apply. It is visible that dilepton data from LHC provides an opportunity to indirectly probe neutrinoless double beta decay, i.e., lepton number violation, specially in the regime of heavy right-handed neutrino masses. We emphasize, the dilepton data itself cannot probe lepton number violation, it is the relation between the $W_R$ and $Z^{\prime}$ masses that results into the bounds shown in Fig.\ \ref{fig:lifetime}. We remind the reader that our findings rely on a flat 70\% signal efficiency,
which is reasonable as shown in Fig.\ 1 of \cite{Aad:2014cka}.

We stress that in the left-right model the $W_R$ mass is determined by the vev of the triplet scalar $\Delta_R$.  Large $W_R$ masses might require the quartic couplings in the scalar potential to be non-perturbative. After including 1-loop effects and renormalization of the scalar sector, the authors in \cite{Maiezza:2016bzp} found that $M_N \leq 7.3 M_{W_R}$ is allowed without ruining stability or perturbativity of the model. 
Thus, the region of parameter space with $M_N > M_{W_R}$ is indeed theoretically allowed. If left-right models with two doublets are considered instead of the triplet scalar a similar logic should apply, concretely reaffirming that for heavy right-handed neutrinos, dilepton data offers a promising search strategy to grasp  left-right symmetry in nature.

As a note we emphasize that the $eejj$ limits obtained in the context of left-right models with no $W-W_R$ mixing are 
equivalent to those right-handed neutrino searches which are parametrized 
in terms of the Lagrangian $\theta \frac{g}{\sqrt{2}} \bar{l}(1 + \gamma_5)N W$, with $\theta=10^{-4}$ being the mixing angle between $W$ and $W_R$ such as in \cite{Deppisch:2015qwa,Chen:2013fna}. 
Thus our conclusions are also applicable to those studies focused on right-handed neutrino searches at the LHC.

\section{Conclusions}

We have performed a collider study of the $Z^{\prime}$ gauge boson in the context of left-right symmetry models motivated by the 
13 TeV dilepton data set with $ 3.2$ fb$^{-1}$ integrated luminosity from the ATLAS collaboration and exploited the complementarity with neutrinoless double beta decay, dijet and $W_R$ searches. Leaving the right-handed gauge coupling free, we set limits of up to 4 TeV on $M_{Z^{\prime}}$, while the limit for the canonical $g_R=g_L$ case is $M_{Z^{\prime}} > 3.2$~TeV. Our findings are nearly independent of the right-handed neutrino masses, as opposed to $eejj$ and double beta decay constraints. 

In the context of minimal left-right models one has $M_{Z^{\prime}}> M_{W_R}$, with the proportionality 
factor of order one depending on the Weinberg angle and the ratio of left-and right-handed gauge couplings. 
This naively indicates that the most promising way to constrain the left-right symmetry 
is by searching for $W_R$ bosons at the LHC. However, as the relation between the gauge 
boson masses depends on model details and can in fact even be 
$M_{Z^{\prime}} \ll M_{W_R}$, $Z^\prime$ searches in the context of left-right 
symmetry should be pursued too. In addition, comparing $W_R$ and $Z^\prime$ properties is an 
important consistency check of possible signals in the future. 
There is yet another motivation for our limits, namely in the context of the complementarity of 
LHC and neutrinoless double beta decay results: if a direct relation between $M_{Z^{\prime}}$ 
and $M_{W_R}$ exists, we have pointed out that 
$Z^{\prime}$ limits constrain the parameter space relevant for double beta decay in an indirect manner, 
reaching areas of parameter space not accessible by dijet and $eejj$ searches 
depending on the ratio of gauge couplings.


\section*{Acknowledgements}
The authors warmly thank Alexandre Alves, Bhupal Dev, Carlos Yaguna, Sudhanwa Patra and Juri Smirnov  for fruitful discussions. WR is supported by the DFG in the Heisenberg Programme with grant RO 2516/6-1. 



\bibliography{darkmatter}

\end{document}